\documentclass[12pt]{article}

\usepackage{url}
\usepackage{xpatch}
\xpatchcmd{\thebibliography}{\section*}{\section*}{}{}

\usepackage{pdfpages}

\usepackage{ragged2e}
\usepackage{amsfonts}
\usepackage{textcomp}

\usepackage{algorithm, algorithmic}
\usepackage{indentfirst}
\usepackage{subfigure}

\usepackage{setspace}

\newlength\savewidth
\newcommand\shline{\noalign{\global\savewidth\arrayrulewidth
                            \global\arrayrulewidth 1pt}
                   \hline
                   \noalign{\global\arrayrulewidth\savewidth}}

\usepackage{amsmath,amssymb}
\usepackage{graphicx}
\usepackage{caption}
\usepackage{color}
\usepackage{multirow}
\usepackage{dcolumn}
\usepackage{bm}
\usepackage{float}
\usepackage{pdfpages}

\definecolor{background-color}{gray}{0.98}
\title{A New Anchor Word Selection Method \\ for the Separable Topic Discovery}

\author{Kun He\thanks{School of Computer Science and Technology, Huazhong University of Science and Technology, Wuhan 430074, China. Corresponding author. Email: brooklet60@hust.edu.cn}, Wu Wang\thanks{School of Computer Science and Technology, Huazhong University of Science and Technology, Wuhan 430074, China} ,  Xiaosen Wang\thanks{School of Computer Science and Technology, Huazhong University of Science and Technology, Wuhan 430074, China.  Corresponding author. Email: xiaosen@hust.edu.cn}, John E. Hopcroft \thanks{Department of Computer Science, Cornell University, Ithaca NY 14853, USA} }

\date{}
\begin{document}
\maketitle

\begin{center}
\subsubsection*{\small Article Type:}
Focus Article

\hfill \break
\thanks

\subsubsection*{Abstract}
\begin{flushleft}
\justifying
Separable Non-negative Matrix Factorization (SNMF) is an important method for topic modeling, where ``separable'' assumes every topic contains at least one anchor word, defined as a word that has non-zero probability only on that topic. SNMF focuses on the word co-occurrence patterns to reveal topics by two steps: anchor word selection and topic recovery. The quality of the anchor words strongly influences the quality of the extracted topics. Existing anchor word selection algorithm is to greedily find an approximate convex hull in a high-dimensional word co-occurrence space. In this work, we propose a new method for the anchor word selection by associating the word co-occurrence probability with the words similarity and assuming that the most different words on semantic are potential candidates for the anchor words. Therefore, if the similarity of a word-pair is very low, then the two words are very likely to be the anchor words. According to the statistical information of text corpora, we can get the similarity of all word-pairs. We build the word similarity graph where the nodes correspond to words and weights on edges stand for the word-pair similarity. Following this way, we design a greedy method to find a minimum edge-weight anchor clique of a given size in the graph for the anchor word selection. Extensive experiments on real-world corpus demonstrate the effectiveness of the proposed anchor word selection method that outperforms the common convex hull-based methods on the revealed topic quality. Meanwhile, our method is much faster than typical SNMF based method.


\end{flushleft}
\end{center}

\clearpage

\renewcommand{\baselinestretch}{1.5}
\normalsize

\clearpage

\section*{\sffamily \Large INTRODUCTION}
Topic modeling is an important problem that aims to discover the abstract topics, a latent thematic structure in a collection of text documents called corpora. The main idea is that documents arise as a distribution on a small number of topic vectors, where each topic vector is a distribution on words. Topic modeling is a crucial task in a wide range of applications, such as content analysis~\cite{jiang2014fast,tong2014tcs}, text classification~\cite{chen2011short}, document clustering~\cite{jin2011transferring} and query suggestion~\cite{jiang2014personalized}. In the field of information retrieval, short text is the dominating content of Internet, such as tweets, news titles and forum messages, and topic modeling plays a key role in mining topics from short texts~\cite{zuo2016topic}. 

Existing topic modeling\footnote{\color{red}Note that our discussion on topic modeling does not consider extensions related to topic modeling, for example the automatic labelling of topic models~\cite{Jey2011topic}. } approaches can be divided into two main categories, probabilistic models~\cite{blei2003latent,hofmann2017probabilistic} and spectral methods~\cite{arora2013practical,arora2012learning,nguyen2014anchors}.

The first category is based on basic probabilistic models, such as Probabilistic Latent Semantic Indexing (PLSI)~\cite{hofmann2017probabilistic} and Latent Dirichlet Allocation (LDA)~\cite{blei2003latent}. LDA model is the most popular and frequently-used topic modeling method. As it is intractable to directly learn the parameters, LDA model uses likelihood-based inference techniques such as Markov Chain Monte Carlo (MCMC)~\cite{griffiths2004finding} and variational inference~\cite{blei2003latent} to learn the parameters. Likelihood-based training requires expensive approximate inference. Gibbs sampling~\cite{li2014reducing,chen2013scalable} and parallel inference~\cite{foulds2013stochastic,yang2014large,liu2015scalable} have been widely studied to improve the performance and scalability.

The second category, spectral method, suggests an algebraic recovery perspective and utilizes Non-negative Matrix Factorization (NMF) as the main technique. Unfortunately, finding exact solution for NMF is NP-hard~\cite{vavasis2009complexity}. Based on the separability assumption, SNMF is able to provide provable polynomial-time algorithms on learning topic models. SNMF for topic modeling consists of two steps, selecting anchor words and recovering a topic matrix, and how to select high quality anchor words plays a key role~\cite{murfi2017accuracy}.

Existing methods for anchor word selection are based on the convex hull searched by either linear programming~\cite{arora2012learning} or combinatorial algorithms~\cite{arora2013practical}. The combinatorial anchor-word-selection method is an improvement over previous work using linear programming. 
As a representative combinatorial method, the \textsc{FastAnchorWords} (FAW) algorithm~\cite{arora2013practical} aims to find an approximate convex hull around a set of vectors in a high-dimensional space. To this end, FAW iteratively finds the farthest point from the subspace spanned by the current set of anchor words until an expected number of anchor words are found.  

Differs to the convex-hull anchor-word-selection method, we associate the word co-occurrence probability with the similarity of the words, and propose \textsc{SoftClique} algorithm (SC). We calculate the similarity of all word-pairs by utilizing the word co-occurrence probability in the corpora. Then, rather than considering words as vectors in the high dimensional space, we regard each word as a node and the word co-occurrence probability as the edge weight to construct the word similarity graph. Based on the separability assumption, we assume the most semantically different words are potential candidates for the anchor words. Therefore, a word-clique with low similarity is likely to be a set of anchor words. Based on this observation, we propose a new anchor word selection algorithm that finds a minimum edge-weight clique of given size $K$ in the word similarity graph. 

We compare our method with the convex hull-based FAW algorithm~\cite{arora2013practical} and the classical LDA method~\cite{blei2003latent} on several real-world corpus with topic coherence.Extensive experiments show that the proposed SC outperforms the baselines. And we compare the running time of SC and FAW on KOS corpora and find SC is faster than FAW. Moreover, we could explore more proporties of topics using SC, such as multiple anchor words for a topic, hidden topic model and so on.

\section*{\sffamily \Large RELATED WORK}
In this section, we provide a review on topic modeling based on NMF and its variant SNMF, which are most related to ourwork.
\subsection*{\sffamily \large Topic Discovery via NMF}
Topic models are statistical models, where each topic is a multinomial distribution over words and a document is represented by a multinomial distribution over topics. Given a corpora $\mathcal{D}$ with $M$ documents, we denote the number of words in the vocabulary by $V$ and the predefined number of topics by $K$. We concatenate the column vectors $\mathbf{A}_{k}~(1 \leq k \leq K)$ for each of the $K$ topics to get the word-topic matrix $\mathbf{A}_{V \times K}$. Similarly, we concatenate the column vectors $\mathbf{W}_{m}~(1 \leq m \leq M)$ for $M$ documents to get the topic-document matrix $\mathbf{W}_{K \times M}$. Given a corpora $\mathcal{D}$, the word-document matrix $\mathbf{X}_{V \times M}$ is the only observed variable where each column corresponds to the empirical word frequencies in the corresponding document. The learning task of topic modeling is to estimate the word-topic matrix $\mathbf{A}$ and the topic-document matrix $\mathbf{W}$, especially the word-topic matrix $\mathbf{A}$. 

Specifically, given a matrix $\mathbf{X}_{V \times M}$ with non-negative entries, NMF will formally decompose the matrix into
\begin{equation}
\mathbf{X} = \mathbf{A}\mathbf{W},
\label{one}
\end{equation}
where $\mathbf{A}$ and $\mathbf{W}$ with small inner-dimension $K$ are non-negative matrices. In general, $K \ll \text{min}\{V,M\}$ is the number of hidden variables, i.e. the number of topics for topic modeling. In practice, NMF faces challenges as it is NP-Hard~\cite{vavasis2009complexity} and highly ill-posed~\cite{gillis2012sparse}. A standard approach is to use the alternating minimization, which is essentially the same as expectation-maximization method (EM) and usually gets stuck in local minima in practice. 

\subsection*{\sffamily \large Topic Discovery via SNMF} 
SNMF is based on the separability assumption~\cite{donoho2004does}, that for each topic, there exists a particular word that only appears in this topic defined as an \textbf{anchor word}.

SNMF for topic modeling begins with the word co-occurrence matrix $\mathbf{Q}$ to avoid the sparseness. The empirical word-word matrix $\mathbf{Q}$ can be understood as a second-order moment matrix of the word-document matrix $\mathbf{X}$:
\begin{equation}
\mathbf{Q} = \mathbf{X}\mathbf{X}^T = \mathbf{A}\mathbf{W}(\mathbf{A}\mathbf{W})^T = \mathbf{A}\mathbf{W}\mathbf{W}^T\mathbf{A}^T = \mathbf{A}\mathbf{R}\mathbf{A}^T,
\label{two}
\end{equation}
where $\mathbf{R} = \mathbf{W}\mathbf{W}^T$. Thus, $\mathbf{Q}_{V \times V}$ is a word-word matrix, $\mathbf{A}_{V \times K}$ is a word-topic matrix, and $\mathbf{R}_{K \times K}$ is a topic-topic matrix. The main goal of topic modeling by SNMF is to recover matrix $\mathbf{A}$ for a given matrix $\mathbf{Q}$.
There are two key steps: 1) find the set of anchor words for the given matrix $\mathbf{Q}$ and 2) recover the word-topic matrix $\mathbf{A}$ with respect to the anchor words. 

\textbf{Algorithm 1} shows the detailed procedure of SNMF. SNMF first finds a set $\mathcal{S} = \{s_1,s_2, ... ,s_K\}$ of $K$ anchor words and subsequently recovers the word-topic matrix $\mathbf{A}$ based on these anchor words. 

\begin{algorithm}[htb]
\caption{\text{SNMF for Topic Discovery}}
\begin{algorithmic}[1]
\REQUIRE corpora $\mathcal{D}$, number of topics $K$
\ENSURE  word-topic matrix $\mathbf{A}$

\STATE Calculate word co-occurrence matrix $\mathbf{Q}$ from corpora $\mathcal{D}$.
\STATE $\mathcal{S} = \text{anchor\_word\_selection}(\mathbf{Q},K)$; // select anchor words as the set $\mathcal{S}$
\STATE $\mathbf{A} = \text{topic\_recovery}(\mathbf{Q},\mathcal{S})$; // recover the word-topic matrix $\mathbf{A}$
\RETURN $\mathcal{S}$
\end{algorithmic}
\end{algorithm}

The current popular method to find anchor words is the convex-hull-based method such as FAW, a purely combinational algorithm that solves the convex hull problem in order to find the anchor words. FAW uses the row normalized version of the word-word matrix $\mathbf{Q}$, denoted as $\mathbf{Q}^\prime$ as the input, each row represents a word in the vocabulary. The rows indexed by anchor words in $\mathcal{S}$ are special. Every other row of $\mathbf{Q}^\prime$ lies in the convex hull of the rows indexed by anchor words. Each row of $\mathbf{Q}^\prime$ is regarded as a vector point in a $V$-dimensional space, so we can obtain $V$ vector points. FAW algorithm first uses random projection via either Gaussian random matrices ~\cite{johnson1984extensions} or sparse random matrices ~\cite{achlioptas2001database} to reduce the dimensionality of each vector point, then it iteratively finds the furthest vector point from the subspace spanned by the set of vector points found so far. The algorithm terminates when $K$ anchor words are found.

At the recovery step, we aim to recover the word-topic matrix $\mathbf{A}$. After we obtain the anchor words, SNMF first needs to find the coefficients matrix that can best reconstruct the matrix $\mathbf{Q}^\prime$ utilizing the rows indexed by the anchor words. Denoting these anchor rows as matrix $\mathbf{Q}^\prime_{S}$, which is a $K \times V$ matrix. We represent the coefficients matrix of the reconstruction as a $V \times K$ matrix $\mathbf{C}$. The reconstruction is:
\vspace{-1.5em}
\begin{equation}
\mathbf{Q}^\prime =\mathbf{C} \mathbf{Q}^\prime_S,
\label{three}
\vspace{-1.5em}
\end{equation}

and matrix $\mathbf{C}$ satisfies the following conditions:
\vspace{-1em}
\begin{equation}
\sum_{k=1}^{K}\mathbf{C}_{i,k} = 1, \mathbf{C}_{i,k} \geq 0.
\vspace{-1em}
\end{equation}
Since all these matrices have practical probability meanings~\cite{arora2013practical}, once we have the coefficients matrix $\mathbf{C}$, we can recover the word-topic matrix $\mathbf{A}$ using Bayes rule. Finally, the word-topic matrix $\mathbf{A}$ is the column normalized version of the matrix $diag(\mathbf{Q}\cdot\vec{\mathbf{1}})\mathbf{C}$.
The reconstruction Eq. (\ref{three}) in the topic recovery step can be decomposed into individual optimization problems, each can be solved using the exponentiated gradient algorithm~\cite{arora2013practical}. We use the same topic recovery subroutine as in FAW in the recovery step , and our contribution is to propose a new method to select high quality anchor words.

\section*{\sffamily \Large THE PROPOSED METHOD}

\begin{figure}[htbp]
\centering
\includegraphics[width=0.90\textwidth]{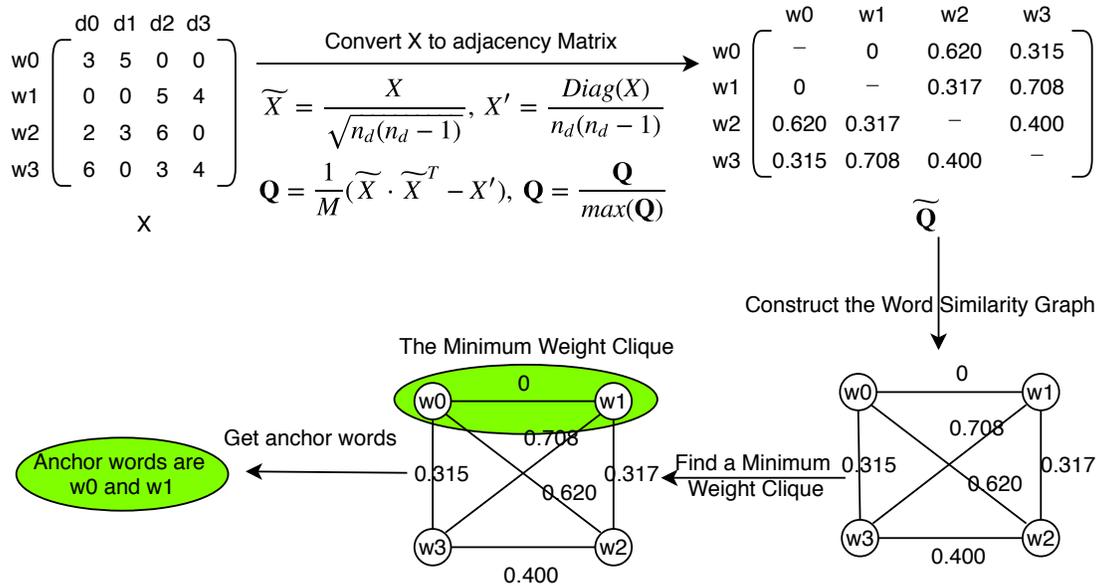}
\caption{ Our method consists two stages, converting $X_{V \times M}$ to the word similarity graph (Subsection of \textit{The Word Similarity Graph}) and finding $K$-cliques (Subsection of \textit{Finding Cliques in the Word Similarity Graph})}
\label{Fig:Example}
\end{figure}

We propose a novel anchor word selection algorithm by finding a minimum edge weight clique in the word similarity graph with matrix $\mathbf{Q}$, which is the second-order moment matrix focused on the word co-occurence pattern. There are two stages in our algorithm, constructing the word similarity graph and finding a minimum edge weight clique of a given size $K$. The total flow of our method is illustrated in Figure \ref{Fig:Example} by a small example. 

\subsection*{\sffamily \large The Word Similarity Graph}
$\mathbf{Q}$ plays a key role in both anchor selection and topic recovery. We can transform $\mathbf{X}$ into $\mathbf{Q}$ to avoid the sparseness of $\mathbf{X}$. How to generate $\mathbf{Q}$ is the same as in Wang et al. ~\cite{wang2016unsupervised}, then we normalize $\mathbf{Q}$ by $\mathbf{Q} = \frac{\mathbf{Q}}{\max(\mathbf{Q})}$. From the probabilistic perspective, each entry of $\mathbf{Q}$ stands for the empirical co-occurrence probability between the two corresponding words. Denote the vocabulary as $\{w_1,w_2,...,w_V\}$, we have $\mathbf{Q}_{i,j}=p(w_i,w_j)$, the $p(w_i,w_j)$ indicates the probability that words $w_i$ and $w_j$ simultaneously appear in the same document. As we know, the conditional probability of two words can express their semantic similarity, emphasizing the associative relations ~\cite{steyvers2007probabilistic}. For example, $p(w_1|w_2)$ is likely that $w_1$ is generated as an associative response to $w_2$.
If the semantic relevance of two words is strong, then the mutual response is more frequent, such as ``NBA'' and ``basketball''.
Noted that the two values $p(w_i|w_j)$ and $p(w_j|w_i)$ are different, because the former reflects the response of $w_i$ to $w_j$, the latter is the opposite. Considering the two values, we utilize the joint probability to express the semantic relevance of a word-pair because the joint probability is: $p(w_i,w_j) \propto p(w_i|w_j), \ p(w_i,w_j) \propto p(w_j|w_i).$

Intuitively, a word-pair with larger word co-occurrence probability is more relevant at the semantic level. Therefore,we can associate the word co-occurrence probability with the semantic relevance. As $\mathbf{Q}$ is symmetric, we can transform it to an undirected weighted graph by regarding $\mathbf{Q}$ as the adjacency matrix, called the word similarity graph.


\subsection*{\sffamily \large Finding Cliques in the Word Similarity Graph}
The separability assumption claims that an anchor word only appears in a specific topic, such that the anchor word strongly indicates the corresponding topic.
As anchor words are a set of words with very weak semantic relevance, we can select anchor words by finding a set of nodes that are connected by very low edge-weights. Specifically, for a given $K$, we can find a $K$-clique with minimum edge-weight in the word similarity graph. Nodes in the clique correspond to the anchor words, and the clique is called anchor clique.

The SC is designed to find the anchor clique. As shown in \textbf{Algorithm 2}, SC includes two steps: \textsc{MergeInit} for greedy initialization and \textsc{LocalSearch} for iterative optimization.
\begin{algorithm}
\caption{\textsc{SoftClique(SC)}}
\begin{algorithmic}[1]
\REQUIRE word co-occurrence matrix $\mathbf{Q}$, number of topics $K$,
\ENSURE anchor word set $\mathcal{S}$
\STATE $MergeInitAnchorSet \leftarrow$ \textsc{MergeInit}($\widetilde{\mathbf{Q}}$, $K$)
\STATE $\mathcal{S} \leftarrow$ \textsc{LocalSearch}($\widetilde{\mathbf{Q}}$, $MergeInitAnchorSet$)
\RETURN $\mathcal{S}$
\end{algorithmic}
\end{algorithm}

In the \textsc{MergeInit} step, the main idea is to initialize the anchor clique with a relatively low-weight clique. We can initialize a clique by greedy expansion from a random node. In practice, we do greedy expansion from multiple nodes, and obtain multiple cliques. From these cliques, we select $K$ nodes that appear the most as the initialized anchor clique. That is, nodes in the intersection of cliques are more likely to be selected.
The \textsc{MergeInit} for greedy initialization is presented in \textbf{Algorithm 3}.

\begin{algorithm}[htb]
\caption{\textsc{MergeInit}}
\begin{algorithmic}[1]
\REQUIRE word co-occurrence matrix $\mathbf{Q}$, number of topics $K$
\ENSURE  initialized anchor clique $\mathcal{S}=\{s_{1},s_{2}, \cdots ,s_{K}\}$

\STATE $V \leftarrow $ number of nodes in $\mathbf{Q}$
\STATE $nodeCnt[V] \leftarrow [0,0...]$
\STATE $startNodes[ ] \leftarrow$ a randomly chosen $\lceil \frac{V}{K} \rceil$ Nodes

\FOR{$startNode$ in $startNodes$}
	\STATE $tmpAnchorSet \leftarrow [startNode]$
	\WHILE{number of words in $tmpAnchorSet < K$}
        \STATE choose node $x$ that minimize $Weight(x,tmpAnchorSet)$
        \STATE $tmpAnchorSet \leftarrow tmpAnchorSet \cup \{x\}$
    \ENDWHILE
    \STATE update corresponding value in $nodeCnt$ by one for nodes in $tmpAnchorSet$
\ENDFOR

\STATE $AnchorSet \leftarrow$ top $K$ words with highest value in $nodeCnt$

\RETURN $AnchorSet$

\textbf{Notation}: $Weight(x,C)$ denotes the weight of node $x$ to clique $C$, namely the total edge-weights from $x$ to each node in clique $C$
\end{algorithmic}
\end{algorithm}

After the greedy initialization, we obtain an initial anchor clique, then we use \textsc{LocalSearch} to update the clique until we obtain a clique that the average edge-weight is very small. The \textsc{LocalSearch} for iterative optimization is presented in \textbf{Algorithm 4}.

\begin{algorithm}[htb]
\caption{\textsc{LocalSearch}}
\begin{algorithmic}[1]
\REQUIRE word co-occurrence matrix $\mathbf{Q}$, initial clique $S$, maximum iteration $max\_iter$
\ENSURE  anchor clique $\mathcal{S}=\{s_{1},s_{2}, \cdots ,s_{K}\}$
\FOR{$i \leftarrow 1$ to $max\_iter$}
    \STATE $wordDiffSet \leftarrow \{\}$
    \FOR{$s$ in $\mathcal{S}$}
        \STATE find word $w$ in $\widetilde{\mathbf{Q}}$ that minimize $val = Weight(w,\mathcal{S} \setminus \{s\})-Weight(s, \mathcal{S}\setminus \{s\})$
        \STATE $wordDiffSet \leftarrow wordDiffSet+(w, s ,val)$
    \ENDFOR
    \STATE $(w', s',val')\leftarrow$ triple with minimum $val'$ in $wordDiffSet$
    \IF{$val' > 0$}
        \STATE \textbf{break}
    \ENDIF
    \STATE $\mathcal{S} \leftarrow (\mathcal{S} \setminus {s'}) \cup \{w'\}$
\ENDFOR
\RETURN $\mathcal{S}$
\end{algorithmic}
\end{algorithm}

\section*{\sffamily \Large EXPERIMENTS}
In this section, we conduct the experiments on real-world corpus, and compare SC with LDA and FAW. All experiments are carried on a Linux server with Intel(R) Xeon(R) 2.00 Ghz CPU and 64G memory. We compare the coherence of topics and running time and further explore the impact of multiple K-cliques.
\subsection*{\sffamily \large Datasets}
We select four real-world corpus of different domains in the UCI Machine Learning Repository ~\cite{frank2010uci} as the datasets, NIPS, KOS, Enron and 20NewsGroup.

The NIPS dataset is derived from NIPS conferences. The KOS corpora is from the Daily KOS blogspot. The Enron corpora is a dataset derived from various emails. The 20NewsGroup corpora is a collection of newsgroup documents. All the corpus need to be preprocessed into the format of bag of words. The details of the corpus are in Table \ref{Table:T1}.

\begin{table}[htbp]
	\begin{center}
		\scalebox{0.9}{
			\begin{tabular}{| c | c  | c  | c |  c |}
				\shline
				\bf{Dataset} & \bf{Domain} & \bf{Doc.}& \bf{Vocab.} & \bf{Dataset Scale} \\
				\shline
				\bf{NIPS}        &Paper       & 1500    & 12419     & small   \\
				\bf{KOS}         &Blog        & 3430    & 6906      & small    \\
				\bf{Enron}       &Email       & 39861   & 28102     & medium    \\
				\bf{20NewsGroup} & News       & 18774   & 61188     & medium    \\
				\shline
		\end{tabular}}
	\end{center}
    \vspace{-1.5em}
	\caption{The statistical information of the four corpus.}
	\vspace{-0.5em}
	\label{Table:T1}
\end{table}

To improve the quality of the corpora, we preprocess the corpora by removing the stop words based on a standard English stopword list\footnote{We use the list of 524 stop words included in the MALLET library.}, low-frequency words with document frequency cutoffs and high-frequency words that appear in more than 80$\%$ of the documents. We set different cutoff value for each corpora. It will give us a new vocabulary, while the number of documents remains the same. Table \ref{Table:T2} shows the suited cutoff value for each corpora and the details after preprocessing. 

\begin{table}[htbp]
	\begin{center}
		\scalebox{0.9}{
			\begin{tabular}{| c | c  | c  | c | c |}
				\shline
				\bf{Dataset} & \bf{Cutoff} & \bf{Vocab.}  & \bf{Avg. Doc. Length} & \bf{Avg. Doc. Scale}\\
				\shline
				\bf{NIPS}        & 50       & 2923     & 963  & Long   \\
				\bf{KOS}         & 60         & 1397     & 95  & Short \\
				\bf{Enron}       & 200       & 3607     & 124  & Short  \\
				\bf{20NewsGroup} & 150        & 1935     & 63  & Short  \\
				\shline
		\end{tabular}}
	\end{center}
    \vspace{-1.5em}
	\caption{The parameter value of cutoff and the statistics of corpus after preprocessing.}
	\vspace{-0.5em}
	\label{Table:T2}
\end{table}
The two baselines, FAW and SC are implemented based on SNMF. We choose LDA as it is a widely used approach for topic modeling. For LDA, we use a fast open-source python implementation of Gibbs sampling\footnote{https://pypi.python.org/pypi/lda} and set the prior parameters $\alpha=0.5$ and $\beta=0.01$. For FAW and SC, we use the RecoverL2 method ~\cite{arora2013practical} at the topic recovery step. In all experiments, the number of topics is fixed $K=100$ and Gibbs sampling is run for 1000 iterations. The final results are the average over five rounds.

\subsection*{\sffamily \large Evaluation of the topic quality}

To intuitively measure the quality of topic models, we choose the topic coherence~\cite{mimno2011optimizing}, which is highly correlated with human judgments of the topic quality.
Topic coherence measures the score of a single topic by computing the semantic similarity between the high probability words in a topic.
Given a topic $z$ and $H$ words that appear most in topic $z$, $W^{(z)}=\{w_{1}, w_{2},\cdots,w_{H}\}$, the topic coherence score is calculated as follows:

\vspace{-1.5em}
\begin{equation}
\textrm{Coherence}(z, W^{(z)}) = \sum_{i=2}^{H} \sum_{j=1}^{i-1} \log \frac{Num(w_{i}, w_{j}) + \epsilon}{Num(w_j)}.
\label{five}
\end{equation}

$Num(w_{i},w_{j})$ is the number of documents that $w_{i}$ and $w_{j}$ appear simultaneously and $Num(w_{i})$ is the number of documents $w_{i}$ appears. $\epsilon$ in the numerator is a smoothing constant to avoid taking the log of zero for words that never co-occur, and we set $\epsilon=10^{-8}$. As words strongly related to the same topic tend to co-occur in the same document, higher topic coherence score implies better topic quality, in other word, better topic interpretability.

To make a general evaluation, we compare the average of the coherence scores of $K = 100$ topics as the evaluation metric for the topic quality, namely,
\begin{equation}
\textrm{Coherence} = \frac{1}{K}\sum_{k=1}^{K}\textrm{Coherence}(z_k, W^{(z_k)}).
\end{equation}

In our experiments, the size of the most probable word set in each topic, ranges from $5$ to $20$, so we set $H = 5,~10,~20$. In the following, we evaluate the performance on each of the four corpus.

The four corpus can be divided into two types, medium corpus (20NewsGroup and Enron) and small corpus (NIPS and KOS). The topic coherence results of the three methods are listed in Table \ref{Table:T3}.

\begin{table}[htbp]
\begin{center}
\scalebox{1.0}{
  \begin{tabular}{| c | c  c c  | c  c  c | c  c  c | c  c  c |}
\shline
\multirow{2}{*}{Algrithm} & \multicolumn{3}{c}{NIPS} \vline& \multicolumn{3}{c|}{KOS} \\
\cline{2-7}
 & $H$~=~5 & $H$~=~10 & $H$~=~20 &\multicolumn{1}{c}{$H$~=~5} & $H$~=~10 & $H$~=~20 \\
\hline
LDA & -193.8     & -866.6    & -3743.2    & -219.1   & -949.1  & -4003.6 \\
FAW &  -134.8 & -599.9    & -2615.6  & -180.0 & -864.2   & -3852.8  \\
SC  & \textbf{-24.5}      & \textbf{-188.9}    & \textbf{-1087.4}      & \textbf{-61.8}   & \textbf{-450.7}  & \textbf{-2509.4} \\
\hline
\multirow{2}{*}{Algrithm} & \multicolumn{3}{c|}{20NewsGroup} & \multicolumn{3}{c|}{Enron} \\
\cline{2-7}
 & $H$~=~5 & $H$~=~10 & $H$~=~20 &\multicolumn{1}{c}{$H$~=~5} & $H$~=~10 & $H$~=~20 \\
 \hline
LDA & \textbf{-187.6}     & -894.8    & -3791.7    & \textbf{-243.4}   & \textbf{-1096.4}  & \textbf{-4642.9} \\
FAW &  -201.7 & -900.0    & -3670.1  & -252.8 & -1137.3   & -4796.9  \\
SC  & -196.6     & \textbf{-874.9}    & \textbf{-3660.6} & -250.5   & -1137.5  & -4823.4 \\
\shline
\end{tabular}}
\end{center}
\vspace{-1.5em}
\caption{Topic coherence on four text corpus. A larger value indicates a set of more coherent topics.}
\vspace{-0.5em}
\label{Table:T3}
\end{table}

On small corpus, we find that SNMF-based methods (FAW and SC) outperform LDA in revealing high-quality topics. For the two SNMF-based methods, SC is clearly much better than FAW. The results indicate SC outperforms FAW and LDA in learning better topics on the two small corpus.

On medium corpora 20NewsGroup, SC is slightly higher than LDA and FAW when $H = 10, 20$, but LDA is the highest when $H = 5$.
On medium corpora Enron, LDA outperforms SC and FAW, and SC is slightly better than FAW. The reason why LDA is better might be that it can use the rich contextual information in a medium set of document collection to learn high-quality topics, while the SNMF-based methods are to find a set of anchor words via heuristics. 

\subsection*{\sffamily \large Running time comparison}
\begin{figure}[htbp]
\centering
\includegraphics[width=0.60\textwidth]{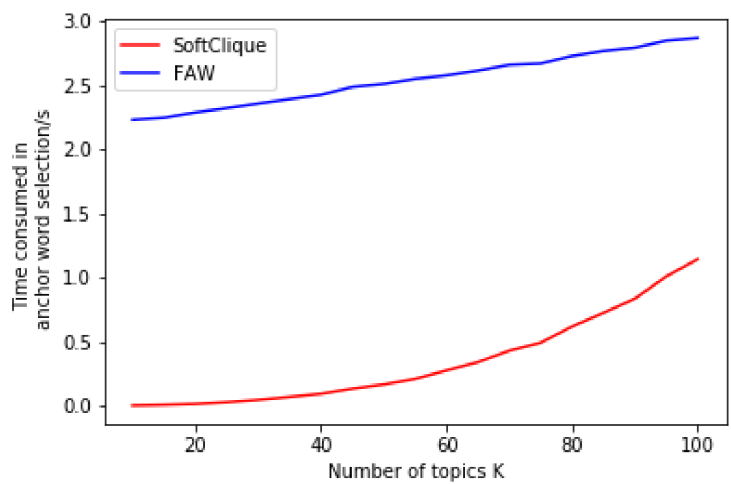}
\caption{The running time of FAW and SC on KOS corpora.}
\label{Fig:Time}
\end{figure}

As for running time, FAW is faster than LDA because it avoids the calculation of linear programming. So we compare the running time of SC with FAW at the anchor word selection step. We conduct experiments on KOS corpora and vary the number of topics from 10 to 100. The result in Figure \ref{Fig:Time} shows that SC is much faster than FAW.

\vspace{-1em}
\subsection*{\sffamily \large Futher exploration on multiple $K$-cliques}
There may be multiple low-weight $K$-cliques in the word similarity graph and we wonder the correlation among these cliques. In this subsection, we do two experiments on NIPS dataset, the cutoff value is listed in Table \ref{Table:T2} and $K$ is set to $5$.

\begin{figure}[htbp]
\centering
\vspace{-3mm}
\scalebox{0.85}{
\includegraphics[width=0.90\textwidth]{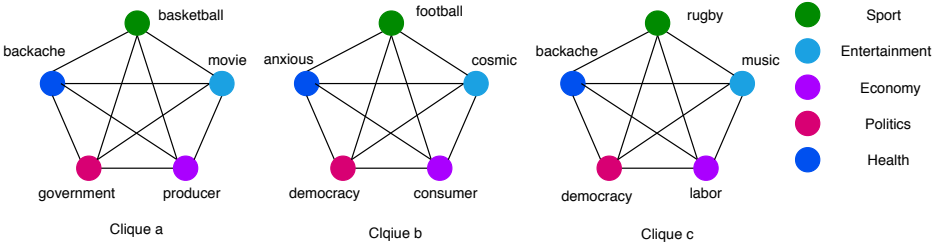}}
\caption{An example to illustrate there could be different anchor words for a topic. We have choosed 5 topics, Sport, Entertainment, Economy, Politics and Health. Clique \textit{a} and Clique \textit{b} don't have any anchor words in common while Clique \textit{c} is mixed with some anchor words from clique \textit{a} and \textit{b}. Nevertheless, each of them could classsify the documents correctly. }
\label{Fig:DiffCliqueWithSameTopic}
\end{figure}

Firstly, we assume that there may be more than one anchor word for a topic. For example, both basketball and football could be the anchor word of sport while we choose only one for each choice.  Figure \ref{Fig:DiffCliqueWithSameTopic} shows a more detailed example. To domenstrate our conjecture, we run 100 rounds to find different anchor words to classify the documents.  By experiments, we got different anchor words and found that some anchor word sets reach similar topics according to the top 10 words in each topic. As an example, we show some anchor words and top 10 words in each topic in Table \ref{Table:T6}. As we see, the top 10 words almost share more than 7 common words in each topic with different anchor words. Therefore, there may be more than one anchor words for each topic and SC could find these anchor words.

\begin{figure}[htbp]
\centering
\vspace{-3mm}
\scalebox{0.7}{
\includegraphics[width=0.90\textwidth]{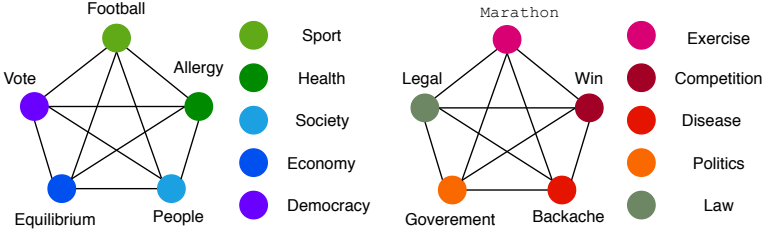}
}
\vspace{-3mm}
\caption{An example to illustrate there may be a hidden topic model apart from topic. There may be five topics of the documents, Sport, Health, Society, Economy and Democracy. While Sport may be divided into competition and exercise and the others could be divided with the same reason. So we may get another five different topics, Exercise, Competition, Disease, Politics and Law. }
\vspace{-1em}
\label{Fig:DiffTopic}
\end{figure}

Secondly, we wonder if there is a hidden topic model apart from the topic model we have found. The hidden topic model means there may be a different classification for the documents. For example, a document about allergy may be classified into either Health or Disease. Figure \ref{Fig:DiffTopic} illustrates a detailed example. To verify this conjecture, we conduct another 100 experiments and find some anchor word sets that classify the documents into different topics with almost the same coherence. Table \ref{Table:T7} lists some anchor word sets. From the top 10 words, we could see that there are at most 3 anchor words in the same topics with almost the same coherence. So, it's not difficult to get the idea that the documents are classified into different topics. Therefore, the hidden topic model does exist and SC could find it.

\begin{spacing}{1.0}
\begin{table}[htbp]
	\begin{center}
		\scalebox{0.8}{
			\begin{tabular}{| p{1.4cm}<{\centering}|  p{5.5cm} | p{5cm} | p{4.5cm} |}
				\shline
				 Anchor words & care, cliff, intensive, likewise, yielding & anatomical, \textbf{chris}, failure, \textbf{firstly}, reader  & \textbf{chris}, \textbf{firstly}, formance, inhibit, preserves\\
				\shline
				Topic 1        & error, neuron, method, \textbf{data}, weight, layer, parameter, \textbf{task}, unit, pattern      & algorithm, weight, neuron, \textbf{information}, error, \textbf{performance}, pattern, method, unit, layer & unit, algorithm, neuron, pattern, method, error, \textbf{hidden}, \textbf{point}, parameter, weight   \\
				\shline
				Topic 2         & unit, output, \textbf{point}, data, training, pattern, \textbf{error}, method, \textbf{space}, vector        & unit, output, data, method, vector, training, parameter, pattern, \textbf{approach}, information     & training, unit, output, pattern, \textbf{weight}, \textbf{signal}, \textbf{information}, \textbf{algorithm}, point, parameter\\
				\shline
				Topic 3         & data, training, \textbf{neuron}, parameter, error, algorithm, point, information, unit, \textbf{part}         & algorithm, parameter, training, \textbf{method}, unit, error, data, point, vector, information     & data, training, \textbf{weight}, parameter, unit, \textbf{signal}, error, algorithm, vector, information\\
				\shline
				Topic 4         & weight, algorithm, training, \textbf{representation}, pattern, unit, method, output, order, vector         & weight, algorithm, \textbf{data}, training, order, unit, \textbf{performance}, error, pattern, term     & output, algorithm, \textbf{neuron}, \textbf{space}, error, vector, weight, method, order, term\\
				\shline
				Topic 5         & algorithm, output, weight, \textbf{information}, data, training, point, \textbf{performance}, part, \textbf{unit}         & output, training, \textbf{neuron}, data, weight, \textbf{order}, error, part, point, term     & algorithm, \textbf{vector}, data, \textbf{part}, error, \textbf{layer}, term, \textbf{representation}, point, map\\
				\shline
		\end{tabular}}
	\end{center}
    \vspace{-1.5em}
	\caption{The top 10 words in each topic with different anchor words but with same topics. We have highlighted the common anchor words. For each topic, we also highlight the unique words.}
	\vspace{-0.5em}
	\label{Table:T6}
\end{table}
\end{spacing}

\begin{spacing}{1.0}
\begin{table}[htbp]
	\begin{center}
		\scalebox{0.8}{
			\begin{tabular}{| p{1.4cm}<{\centering}|  p{5.5cm} | p{5cm} | p{4.5cm} |}
				\shline
				 Anchor words & ble, confidence, hasselmo, nonlinearities, topographic & emergent, horizon, performances, query, summing  & accordance, curvature, molecular, reward, sake\\
				\shline
				Topic 1        & \textbf{neuron}, unit, \textbf{information}, data, distribution, object, examples, \textbf{error}, equation, processing     & layer, algorithm, weight, \textbf{neuron}, \textbf{information}, vector, \textbf{error}, point, processing, linear & \textbf{error}, algorithm, training, unit, \textbf{information}, output, signal, linear, \textbf{neuron}, parameter   \\
				\shline
				Topic 2         & algorithm, unit, part, output, pattern, parameter, point, distribution, \textbf{error}, weight        & output, algorithm, dynamic, space, case, \textbf{error}, simple, term, level, neuron    & data, \textbf{error}, pattern, vector, point, values, space, case, noise, unit\\
				\shline
				Topic 3         & data, weight, linear, \textbf{method}, training, single, error, performance, part, \textbf{vector}         & algorithm, \textbf{method}, unit, data, \textbf{vector}, parameter, space, control, term, signal     & algorithm, training, weight, unit, \textbf{method}, approach, layer, parameter, \textbf{vector}, signal\\
				\shline
				Topic 4         & \textbf{output}, textbf{data}, algorithm, training, method, neuron, \textbf{information}, unit, weight, case         & training, textbf{data}, \textbf{output}, unit, weight, \textbf{information}, part, error, local, values     & textbf{data}, distribution, \textbf{output}, method, performance, \textbf{information}, local, neuron, point, case\\
				\shline
				Topic 5         & training, pattern, weight, parameter, component, point, dynamic, error, probability, equation        & weight, algorithm, unit, training, pattern, data, error, parameter, method, neuron     & point, neuron, output, method, vector, matrix, space, hidden, step, term\\
				\shline
		\end{tabular}}
	\end{center}
    \vspace{-1.5em}
	\caption{The top 10 words in each topic with different words and different topics with coherence between -40 to -30. We have highlighted the common words that appears in all three anchor words result for the same topic.}
	\vspace{-0.5em}
	\label{Table:T7}
\end{table}
\end{spacing}
\vspace{-0.5em}
\section*{\sffamily \Large CONCLUSIONS AND FUTURE WORK}
How to select a more accurate set of anchor words is an open problem for topic mideling. Using a word co-occurrence matrix, we build a word similarity graph that regards words as nodes and co-occurrence probability as edge weights. Further, we associate the word co-occurrence probability with the similarity of words and assume that the most different words on the semantic are the potential candidates for the anchor words. From a new perspective, we propose a new anchor selection method by finding a $K$-clique with the minimum edge-weight in the word similarity graph. Experimental results on real-world corpus suggest that our method outperforms FAW and LDA with Gibbs sampling for the topic quality in small corpus, but the topic interpretability is slightly lower than LDA in medium scale corpus. Besides, SC could find different anchor word sets with almost the same coherence and they are both meaningful.

In future work, we will study promising methods for better anchor selection on large scale corpus. In this paper, we apply a simple local search method with greedy initialization to find a minimum edge-weight clique of size $K$ in the word similarity graph. There exists promising space for potential improvement utilizing more sophisticated algorithms in graph theory. Also, our method probably does not require the anchor words with the minimum edge-weight for the anchor word selection. We will conduct other combinatorial methods in finding the anchor words from the word similarity graph.
Moreover, there may be no need to provide $K$ to finish the anchor word selection with SC method because we could find a clique with proper size so as to find a set of anchor words, which is different from existing methods. As far as we know, there exists little work for automatically determining $K$, so it is difficult to find a criterion to evaluate the quality of $K$. In future, we will try to evlauate the quality of $K$, such as KL divergency of the probability of a document belonging to a topic, the perplexity of the topic model to help determine $K$ automatically.
\vspace{-1em}
\section*{\sffamily \Large ACKNOWLEDGMENTS}
Supported by National Natural Science Foundation of China (61772219, 61472147).

\bibliographystyle{unsrt}

\end{document}